\documentclass[11pt,a4paper]{article}
\usepackage{jheppub}
\title{Why one can maintain that there is a probability loophole in the CHSH}
\author[a]{Han Geurdes,}
\affiliation[a]{Institution,\\
C. vd Lijnstraat 164 2593 NN Den Haag, Netherlands}
\emailAdd{han.geurdes@gmail.com}
\abstract{
In the paper it is demonstrated that the particular form of CHSH, $S=E\{A(1)[B(1)-B(2)]-A(2)[B(1)+B(2)]\}$ with, $S$ maximally $2$ and minimally$-2$,  for $A$ and $B$ functions $\in \{-1,1\}$, is not generally valid. The nonzero probability that local hidden extra parameters  violate the CHSH, is not eliminated with basic principles derived from the CHSH. }
\begin{document}
\maketitle

\section{Introduction and test of the CHSH}
The CHSH inequality is an element in the discussion about the existence or nonexistence of additional local hidden parameters \cite{Eins}. The CHSH inequality \cite{Clauser} is derived from Bells formula for the correlation \cite{Bell}, $E(a,b)$, between distant spin measurements with setting setting parameters $a$ and $b$. Generally,
\begin{equation}\label{1}
E(a,b)=\int d\lambda \rho_{\lambda}A_{\lambda}(a)B_{\lambda}(b)
\end{equation}
In (\ref{1}) we can identify the probability density $\rho_{\lambda} \geq 0$, with $\int d \lambda \rho_{\lambda}=1$. The $\lambda$ are introduced to explain the correlation and need to have a local effect. This can e.g. be accomplished \cite{Geurdes} if a $\lambda_1$ is assigned to the $A$ measurement instrument and $\lambda_2$ to the $B$ instrument. Furthermore, the measurement functions $A_{\lambda}(a)$ and $B_{\lambda}(b)$ both project in $\{-1,1\}$ to represent binairy spin variables (e.g. up=1, down=-1).  The CHSH inequality is based on the following expression,
\begin{equation}\label{2}
S=E(1,1)-E(1,2)-E(2,1)-E(2,2)
\end{equation}
The quartet of setting pairs $\mathcal{Q}=\{(1,1),(1,2),(2,1),(2,2)\}$ occurs random in a series of $N$ spin measurements of entangled particle pairs. Alice and Bob are two assitents in the experiment who, per trial or particle pair measurement, randomly select the setting of their measurement instrument. The argument in favor of the CHSH inequality \cite{Gill} and against a possible probability loophole \cite{Geurdes} is as follows. From (\ref{1}) and (\ref{2}) we may write, suppressing the hidden variables index $\lambda$, notation for the moment, 
\begin{equation}\label{2a}
S=E\{A(1)[B(1)-B(2)]-A(2)[B(1)+B(2)]\}.
\end{equation}
According to \cite{Gill}, because, $A$ and $B$ are both $\in \{-1,1\}$, we see that when $B(1)=B(2)$, then $S=\pm 2$, while, when $B(1)=-B(2)$, it again flollows, $S=\pm 2$. Hence, $|S|$ based on (\ref{1}) cannot be larger than $2$ and therefore the nonzero probability of $|S|>2$ with a local hidden variables model of \cite{Geurdes} must be based on a mistake. It will be demonstrated in the next section that this claim is untrue. In the paper we show that this argument does not hold in general. The loophole paper \cite{Geurdes} has the intention to derive a test of the strength of conclusions that can be derived from the CHSH inequality. Tests of strength are not uncommon in statistcs. In \cite{Geurdes} this is done via a reformulation of Bells formula. Let us define sets based on the difference $E(a,b)-E(x,y)$, $(a,b)$ and $(x,y)$ are different settings. We have, $\Omega_{+}(a,b;x,y)=\{\lambda \,|\, A_{\lambda}(a)B_{\lambda}(b) = A_{\lambda}(x)B_{\lambda}(y) =+1\}$, together with $\Omega_{-}(a,b;x,y)=\{\lambda \,|\, A_{\lambda}(a)B_{\lambda}(b) = A_{\lambda}(x)B_{\lambda}(y) =-1\}$ and $\Omega_{0}(a,b;x,y)=\{\lambda \,|\, A_{\lambda}(a)B_{\lambda}(b) = - A_{\lambda}(x)B_{\lambda}(y) =\pm 1\}$. The three sets are disjoint and if $\Lambda$ denotes the universe set of the $\lambda$ variables we also have $\Lambda=\Omega_{+}(a,b;x,y) \cup \Omega_{-}(a,b;x,y) \cup \Omega_{0}(a,b;x,y)$. Note that in $E(a,b)-E(x,y)$ only the $\lambda \in \Omega_{0}(a,b;x,y)$ contribute. Therefore
\begin{equation}\label{3}
E(a,b)-E(x,y)=-2\int_{\lambda \in \Omega_{0}(a,b;x,y)} A_{\lambda}(x)B_{\lambda}(y) d\lambda
\end{equation}
If subsequently, $E(a,b)=0$ and we write $\Omega^{\prime}_0(x,y)= \Omega_{0}(a,b;x,y) $ and $(a,b)$ such that $E(a,b)=0$, then
\begin{equation}\label{4}
E(x,y)= 2\int_{\lambda \in \Omega^{\prime}_{0}(x,y)}\rho_{\lambda} A_{\lambda}(x)B_{\lambda}(y) d\lambda
\end{equation}
With $E(x,y)=E_T(x,y)$. Subsequently from $E(a,b)=0$ it follows \cite{Geurdes} that,
\begin{equation}\label{5}
E_C(x,y)=2\int_{\lambda \in \Omega^{\prime}_{+}(x,y)}\rho_{\lambda}d \lambda -2\int_{\lambda \in \Omega^{\prime}_{-}(x,y)}\rho_{\lambda}d \lambda
\end{equation}
and, of course, $E_C(x,y)=E(x,y)$ via $E(a,b)=0$.

The settings that we employ are, for Alice, $1_A=(1,0,0)^T$ and $2_A=(0,0,1)^T$. The superscript $T$ means transposed of a vector. For Bob we take, $1_B=\frac{1}{\sqrt{2}}(1,1,0)^T$ and $2_B=\frac{1}{\sqrt{2}}(-1,0,-1)^T$. If the $A$ and $B$ indices in $1_A$ etc, are not necessary they will be omitted. With this selection of setting vectors and taking the quantum correlation innerproduct $\langle a, b\rangle =\sum_{i=1}^3 a_ib_i$, the S from (\ref{2}) will produce $|S|=\frac{3}{\sqrt{2}} > 2 $. Like in \cite{Geurdes} we take the probability density, $\rho_{\lambda}=\rho_{\lambda_1,\lambda_2}$ and $\rho_{\lambda_1,\lambda_2}=\rho_{\lambda_1}\rho_{\lambda_2}$. The separate $\lambda_1$, is assigned to $A$ and $\lambda_2$, is assigned to $B$. For, $j=1,2$,
\begin{eqnarray}\label{6}
\rho_{\lambda_{j}}=\{
\begin{array}{ll}
\frac{1}{\sqrt{2}},\, \lambda_j \in \left[-\frac{1}{\sqrt{2}},\frac{1}{\sqrt{2}}\right]=\Lambda_j \\
0,~~~ \lambda_j \notin \left[-\frac{1}{\sqrt{2}},\frac{1}{\sqrt{2}}\right]\\
\end{array}
\end{eqnarray}
with the universal set, $\Lambda=\Lambda_1 \times \Lambda_2$. Furthermore, $\Omega^{\prime}_{\pm}(x,y)$, is the Cartesian product of a $\lambda_1$ and a $\lambda_2$ interval, i.e. $ \Omega^{\prime}_{\pm}(x,y) =\Omega^{\prime}_{A \pm}(x) \times \Omega^{\prime}_{B \pm}(y)$. Similar as in \cite{Geurdes} let us take
\begin{eqnarray}\label{7}
\begin{array}{ll}
\Omega^{\prime}_{A \pm}(1)\, \in \left \{\emptyset, \{ \lambda_1 \,|\, -1 + \frac{1}{\sqrt{2}} \leq \lambda_1 \leq \frac{1}{\sqrt{2}}\} \right\},~\Omega^{\prime}_{B \pm}(1)\, \in \left \{\emptyset, \{ \lambda_2 \,|\, -\frac{1}{\sqrt{2}} \leq \lambda_2 \leq 0 \} \right\}
\\ 
\Omega^{\prime}_{A \pm}(2)\, \in \left\{\emptyset, \{ \lambda_1 \,|\, - \frac{1}{\sqrt{2}} \leq \lambda_1 \leq 1-  \frac{1}{\sqrt{2}} \}  \right\},~
\Omega^{\prime}_{B \pm}(2)\, \in \left\{\emptyset, \{ \lambda_2 \,|\,  0< \lambda_2 \leq \frac{1}{\sqrt{2}} \}  \right\}\\ 
\end{array}
\end{eqnarray}
Note, $\int_{\lambda_j \in \emptyset} \rho_{\lambda_j}d\lambda_j=0$. The following form will be used in the study of (\ref{2a}),
\begin{equation}\label{8}
E_C(x,y)=\int_{\Omega^{\prime}_{A +}(x)}d \lambda_1\int_{\Omega^{\prime}_{B +}(y)}d \lambda_2 - \int_{\Omega^{\prime}_{A -}(x)}d \lambda_1\int_{\Omega^{\prime}_{B -}(y)}d \lambda_2
\end{equation}
In order to have a similar approach as in (\ref{2a}) we introduce $\theta_{\cdot,\cdot}^{\cdot}(x)$ forms will be defined from the sets in (\ref{7}). For instance let us define
\begin{eqnarray}\label{9}
\theta_{A\, \lambda_1}^{\pm}(x)=\{
\begin{array}{ll}
1,~~\lambda_1 \in \Omega^{\prime}_{A \pm}(x)\neq \emptyset,\\
0,~~\lambda_1 \notin \Omega^{\prime}_{A \pm}(x).\\
\end{array}
\end{eqnarray}
We note that $\theta_{A\, \lambda_1}^{\pm}(x)=0$, when, $\Omega^{\prime}_{A \pm}(x)=\emptyset$. Similarly, 
\begin{eqnarray}\label{10}
\theta_{B\, \lambda_2}^{\pm}(y)=\{
\begin{array}{ll}
1,~~\lambda_2 \in \Omega^{\prime}_{B \pm}(y)\neq \emptyset,\\
0,~~\lambda_2 \notin \Omega^{\prime}_{B \pm}(y).\\
\end{array}
\end{eqnarray}
and $\theta_{B\, \lambda_2}^{\pm}(y)=0$ when $\Omega^{\prime}_{B \pm}(y)=\emptyset$. Given the expressions in (\ref{8}) - (\ref{10}), the $S$ given in (\ref{2a}) can be similarly given by
\begin{eqnarray}\label{11}
\begin{array}{ll}
S=\int_{\Lambda_1} d\lambda_1 \int_{\Lambda_2} d \lambda_2 [
\theta_{A\, \lambda_1}^{+}(1)\theta_{B\, \lambda_2}^{+}(1)- \theta_{A\, \lambda_1}^{-}(1)\theta_{B\, \lambda_2}^{-}(1) \\
\hspace{1.3 in} -\theta_{A\, \lambda_1}^{+}(1)\theta_{B\, \lambda_2}^{+}(2) + \theta_{A\, \lambda_1}^{-}(1)\theta_{B\, \lambda_2}^{-}(2) \\
\hspace{1.3 in}-\theta_{A\, \lambda_1}^{+}(2)\theta_{B\, \lambda_2}^{+}(1) + \theta_{A\, \lambda_1}^{-}(2)\theta_{B\, \lambda_2}^{-}(1) \\
\hspace{1.3 in}-\theta_{A\, \lambda_1}^{+}(2)\theta_{B\, \lambda_2}^{+}(2) + \theta_{A\, \lambda_1}^{-}(2)\theta_{B\, \lambda_2}^{-}(2) ]\\
\end{array}
\end{eqnarray}
 Suppose we take the following values for the $\theta_{\cdot,\cdot}^{\cdot}\in \{0,1\}$ variables in (\ref{11}).
\begin{eqnarray}\label{12}
\begin{array}{ll}
\theta_{A\, \lambda_1}^{+}(1)=\theta_{B\, \lambda_2}^{+}(1)=1,
\theta_{A\, \lambda_1}^{-}(1)=1,
\theta_{B\, \lambda_2}^{-}(1)=0,\\
\theta_{A\, \lambda_1}^{-}(2)=\theta_{B\, \lambda_2}^{-}(2)=1,
\theta_{A\, \lambda_1}^{+}(2)=\theta_{B\, \lambda_2}^{+}(2)=0.\\
\end{array}
\end{eqnarray}
The possibility of selection of $\theta_{\cdot,\cdot}^{\cdot}\in \{0,1\}$  such as in (\ref{12}) cannot be rejected. With this selection of $\theta_{\cdot,\cdot}^{\cdot}\in \{0,1\}$ variables, possible confusion of "multiple random models" is avoided. The averaging over models $\mathcal{L}$ such as was done in \cite{Gill} does not apply to the present case. Its use in \cite{Gill} was already questionable. In \cite{Geurdes} there is only one single fixed model with random input. In the present paper the line of reasoning presented in (\ref{2a}) which was also used in \cite{Gill} to reject the conclusions from \cite{Geurdes} leads us to 
\begin{equation}\label{13}
S=\int_{-1+\frac{1}{\sqrt{2}}}^{+\frac{1}{\sqrt{2}}}d\lambda_1\int_{-\frac{1}{\sqrt{2}}}^0d\lambda_2+\int_{-1+\frac{1}{\sqrt{2}}}^{+\frac{1}{\sqrt{2}}}d\lambda_1\int_{0}^{\frac{1}{\sqrt{2}}}\lambda_2+\int_{-\frac{1}{\sqrt{2}}}^{1-\frac{1}{\sqrt{2}}}d\lambda_1\int^{\frac{1}{\sqrt{2}}}_0d\lambda_2
\end{equation}
Hence, $S=\frac{3}{\sqrt{2}}$ and therefore $|S|>2$ with a single fixed local hidden variables model where the method of deriving $S$ is similar to the way it is used in \cite{Gill}. 

\section{Conclusion \& Discussion}
We conclude that the conjecture in \cite{Gill} that "there must be a mistake in \cite{Geurdes}" is unjustified and that indeed local models can violate the CHSH criterion with nonzero probability. A possible objection that $A$ and $B$ functions do not exist is unfounded.   Both $E_C$ as well as $E_T$ are equivalent to the same Bell formula. Conclusions for $E_C$ are valid for $E_T$ and vice versa. In this paper it was demonstrated that the step from term-by-term: $S=E(1,1)-E(1,2)-E(2,1)-E(2,2)$, to compact $S=E\{A(1)[B(1)-B(2)]-A(2)[B(1)+B(2)]\}$ and therefore $S$ for $A$ and $B$ both $\in \{-1,1\}$ varying between and including $-2$ and $2$, does not hold in all cases. If the step from term-by-term to compact is allowed in the original expression of Bells formula then it is allowed in the $\Omega$ set analysis of Bells formula in \cite{Geurdes}. We showed that a single fixed model of local hidden variables may violate CHSH bounds. Therefore, local hidden variables in e.g. tHoofts predeterminism \cite{Hooft} or in mirror matter \cite{Okun},  \cite{Footb}, \cite{Ciarcell}, \cite{Lee}, \cite{Wu} are still a possibility for the explanation of the entanglement correlation. Perhaps there is no explanation beyond randomness. It is reasonable to expect that this conclusion is arrived at with properly tested statistics.

\end{document}